\documentclass[aps,showpacs]{revtex4}

\usepackage{graphicx}

\begin{document}

\author{F.\ Marty Ytreberg}
\affiliation{Department of Physics, Whitman College, Walla Walla,
  WA 99362}
\author{Susan R.\ McKay}
\affiliation{Department of Physics and Astronomy, The U.\ of Maine, Orono,
  ME 04469-5709}
\date{\today}
\title{Scaling Behavior of Field-Induced Aggregates in Ferrofluids}

\begin{abstract}
An ordered hexagonal array of aggregates can form in thin
ferrofluid layers when an external magnetic field is applied.
Using the Helmholtz free energy for this system, we calculate
the optimum spacing for these aggregates. Results show excellent
agreement with experimental findings as a function of field
strength and layer thickness. Our analysis yields a crossover in
the exponent for the scaling behavior of the aggregate spacing
as a function of plate separation, in agreement with experiment.
For the first time, we report a similar crossover
in the scaling behavior of the aggregate spacing as a function
of the magnetic field. The mechanisms responsible for both
crossovers are introduced and discussed.
\end{abstract}


\maketitle

\section{introduction}

Ferrofluids are composed of small single-domain magnetic particles
dissolved in a liquid carrier. With no external
magnetic field, Brownian motion keeps the particles homogeneously
distributed throughout the liquid. When a sample of ferrofluid is
confined between two glass plates
with a separation $L$, and an external magnetic field $H_0$ is
applied normal to the plates, the particles aggregate.
These aggregates are of approximately uniform size, and
the separation of the aggregates is periodic with an average spacing
$d$. A view normal to the plates displays a hexagonal pattern
of aggregates \cite{hong97,blums96,rosen85}.

In this paper we present a new model for predicting the aggregate
spacing, and compare the results of this model to experiment. We
demonstrate that this model shows good agreement with experimental
data for the aggregate spacing as a function of field $H_0$ and
plate separation $L$. We compare our results to experimental data
from Refs.\ \cite{hong97,wang94}.

The model also yields the scaling behavior of the aggregate
spacing $d$ as a function of $H_0$ and $L$. In 1994, Wang {\it et
al.}\ \cite{wang94} showed that there was a crossover in the
exponent for $d$ as a function of $L$. They attributed this
crossover to an experimentally observed structural change in the
aggregates. In 1995, Liu {\it et al.}\ \cite{liu95} also showed a
crossover in this exponent for ferrofluid emulsions, but no
explanation for the crossover was provided. In our model, this
crossover takes place, and is not attributed to the mechanism in
Ref.\ \cite{wang94}. We also report, for the first time, a
crossover in the scaling behavior for $d$ as a function of $H_0$.
The mechanisms responsible for both crossovers in our model are
identified and discussed.

\section{Helmholtz free energy}

Experiments show that, when the field is first turned on, the
particles form single-particle-diameter chains; the system then
evolves by the chains sticking together to form aggregates. The
time scale associated with the chain formation is much smaller
than the time scale observed for the formation of aggregates from
chains \cite{hong97}. Due to this observation, we treat the
formation of the aggregates as a quasiequilibrium process. Other
examples of this treatment can be seen in Refs.\
\cite{liu95,wang94,zubarev97}. With this assumption, the
separation of the aggregates can be determined by minimizing the
Helmholtz free energy for the ferrofluid system.

There are three contributions to the Helmholtz free energy that must be
considered: the magnetic energy, the entropy
and the surface energy. For the purpose of notational simplicity,
we introduce the constants $\alpha$ and $\beta$, such that
$$
  d=\beta b,\;\;\beta\equiv\sqrt{\frac{2\pi\gamma}{\phi\sqrt{3}}},
$$
$$
  N_0=\frac{\alpha}{b^2}=\frac{\alpha\beta^2}{d^2},
  \;\;\alpha\equiv\frac{\phi \ell^2}{\gamma\pi},
$$
where $N_0$ is the total number of aggregates in the system.
Here, $b$ is the aggregate radius, $\phi$ is the volume fraction of the
ferrofluid, $\gamma$ is the packing fraction of the particles in an
aggregate and $\ell^2$ is the area of one glass plate. These relationships
are purely geometric, using the assumption that the aggregates
are of uniform size and spacing.

\subsection{Magnetic Energy}

The magnetic energy contains three parts corresponding to the
self-energy of the aggregates, the aggregate-aggregate
interactions, and the aggregate-external field interactions. The
model approximates the aggregates as cylinders that are uniformly
magnetized.

The approximation of cylindrical aggregates is consistent with the
experiments in Refs.\ \cite{liu95,ivey00}, where it was found that
a few particles away from the end of the aggregate the shape was
cylindrical. Since we are dealing with $L \sim 10 \, \mu $m and $b
\sim 5$ nm, the length of an aggregate is on the order of a
thousand particles. Given these experimental features, the
approximation of cylindrical aggregates is a good one.

The approximation that the magnetization is uniform throughout the
aggregates is reasonable. Near the ends of the aggregates one
would expect the average magnetic moment of the particles to be
different than that of the bulk. However, the magnetization of the
aggregates is determined mainly by the particles in the bulk. A
correction for surface effects is included in the surface energy,
discussed below. The uniform magnetization approximation has also
been used in Refs.\ \cite{liu95,wang94}.

From magnetostatics we know that the effective surface and volume
charge densities are $\sigma_m=\vec{M}\cdot\hat{n}$ and
$\rho_m=\vec{\nabla}\cdot\vec{M}$ respectively, where $\hat{n}$ is
the unit outward normal, and $\vec{M}$ represents the
magnetization of an aggregate. For the case of uniform
magnetization parallel to the external magnetic field,
$\sigma_m=M$ on the top and bottom of the aggregates, $\sigma_m=0$
on the sides and $\rho_m=0$ everywhere. Thus, the total magnetic
energy of the system can be calculated as interactions between
disks of uniform equivalent charge density $\sigma_m=M$.
\cite{jackson}

For the first contribution to the magnetic energy we consider the
energy required to accumulate a single disk of equivalent charge.
This self-energy of the disk can be calculated exactly in the
following simple way. Consider a disk of radius $r \ll b$ and
uniform charge density $\sigma_m$. The potential at a point on the
rim of the disk is
$$
  V(r,\sigma_m)=4\sigma_m r.
$$
Now bring in a ring of uniform charge density $\sigma_m$, radius
$r$ and thickness $dr$. The additional self-energy of the disk
when the ring is added is
$$
  dE_{self_1}=V(r,\sigma_m)dq=8\pi\sigma_m^2 r^2dr.
$$
Integrating from $r=0$ to $r=b$ yields the total self-energy of
one disk,
$$
  E_{self_1}=\frac{8}{3}\pi\sigma_m^2 b^3.
$$
To obtain the self-energy for all of the disks in the system, we
must multiply this expression by $2N_0$ and use the fact that
$\sigma_m=M$. The factor of two is due to the fact that each
aggregate is replaced by two disks of equivalent charge, so there
are two layers of $N_0$ disks to consider. The final expression
for the self-energy of the system is then
$$
  E_{self}=\frac{16}{3}\pi\alpha M^2 \frac{d}{\beta}.
  \label{EQ:selfE}
$$
This energy partially captures the particle-particle interactions
within an aggregate. An additional contribution to the single
aggregate energy arises from the interaction between the pair of
disks associated with the same aggregate, as discussed below.

For the other contributions to the magnetic energy, the disks can
be treated as point charges. This approximation is valid since the
aggregate spacing and length are large when compared to the
aggregate diameter (i.e.\ $d \gg b$ and $L \gg b$). The condition
that $d \gg b$ is satisfied by having a low volume fraction
$\phi$. (The experiments used for comparison with our model are
done for $\phi < 0.20$.) To satisfy the second condition, we use
experimental observations which indicate that larger values of $L$
produce a smaller $b/L$ ratio. Once the energetically favorable
radii have been determined by minimizing the free energy, points
where $b/L \ll 1$ are considered accurate and other data points
where $b/L \sim 1$ are discarded.

Using the point charge approximation, the aggregates behave as
dipoles, and the interaction energy of the aggregates with the
external magnetic field is
\begin{equation}
  E_{\rm H_0}=N_0(-\vec{m} \cdot \vec{H_0})=-\alpha\pi LMH_0,
\end{equation}
where $\vec{m}$ is the magnetic moment of an aggregate.
This term does not depend on $d$, and therefore will be
considered a constant in the minimization process.

The last contribution to the magnetic energy is due to the
interactions between the disks. We must account for interactions
between disks in the same plane ($E_{\rm same}$) and disks in
opposite planes. For the interaction of the disks on opposite
planes, two terms must be included. One is the interaction of
disks due to the same aggregate ($E_{\rm opp}$, i.e.\ one disk
directly over the other disk), and the other is the interaction of
disks positioned diagonally to each other ($E_{\rm diag}$). Using
the point charge approximation, as previously mentioned, the total
disk interaction energy will be
$$
  E_{\rm inter}=E_{\rm opp}+E_{\rm diag}+E_{\rm same}
  =\alpha\sigma_m^2\pi^2\biggl(-\frac{d^2}{\beta^2L}
   -6\frac{d^2}{\beta^2\sqrt{d^2+L^2}}
   +6\frac{d}{\beta^2}\biggr).
$$
It should be noted that $E_{\rm opp}$ arises from
particle-particle interactions in a single aggregate, while
$E_{\rm same}$ and $E_{\rm diag}$ arise from the interaction
between two separate aggregates.

In Ref.\ \cite{hong97}, Hong and collaborators found that the
magnetization of their ferrofluid sample was well approximated by
to the Langevin function. Since this functional form for the
magnetization is also in agreement with Refs.\
\cite{blums96,rosen85}, and for small $\phi$ in Ref.\
\cite{huke00}, we have used it for our model. Combining these
contributions yields a total magnetic energy of
\begin{equation}
   E_{\rm mag}=\Biggl[ \Biggl(\frac{16}{3\pi}+\frac{6}{\beta}\Biggr)
     \frac{d}{\beta}-\frac{d^2}{\beta^2L}-\frac{6d^2}{\beta^2\sqrt{d^2+L^2}}
     \Biggr]\alpha\pi^2M^2+{\rm const},
  \label{EQ:magE}
\end{equation}
where
\begin{equation}
  M=\frac{\phi}{\gamma}M_S\Biggl( {\rm coth}\eta-\frac{1}{\eta} \Biggr),
    \;\;\eta\equiv\frac{\mu}{k_BT}H_0.
  \label{EQ:M}
\end{equation}
Here, $\mu$ is the magnetic moment of the particles in the ferrofluid and
$M_S$ is the saturation magnetization of the fluid. In the case
of a ferrofluid emulsion, each drop has a magnetization described by
the Langevin function. In this case, the magnetization of an aggregate is
still given by Eq.\ (\ref{EQ:M}), with an important reminder that $\mu$ is
the magnetic moment of the particles in the drop, not the magnetic moment
of the drop itself.

Performing an expansion of Eq.\ (\ref{EQ:magE}) in powers of $d/L$
(with $d=\beta b$), and keeping up to quadratic terms, this
magnetic energy reduces to the one used in Refs.\
\cite{liu95,ytreberg00}. However, one should note that the
assumption $d \ll L$ is a much stronger assumption than $b \ll L$.
For very small volume fractions, one can have the situation where
$d \sim L$ with the condition $b \ll L$ easily satisfied, since
$\beta$ increases as the volume fraction decreases.

This magnetic energy offers a significant advantage over the model
presented in Ref.\ \cite{ytreberg00}, due to the elimination of
all free parameters in the magnetic energy. With no free
parameters in the magnetic energy, we can compare the orders of
magnitude of the magnetic energy and entropy, as discussed in the
next section.

\subsection{Entropy and Surface Energy}

To calculate the entropy we treat the aggregates as
distinguishable. The number of states accessible to the system is
$N_0!$, and the configurational entropy of the system is $S=k_B
\ln{(N_0!)}$. For the experimental systems that we are modelling,
$E_{\rm mag}/TS \sim 10^{4}$, where $T$ is taken to be room
temperature. Thus, the entropy is negligible for this particular
set of experimental conditions.

The last contribution to the energy to be considered is the
surface energy. We have already accounted for particle-particle
interactions in the bulk in calculating the magnetic energy. The
surface energy allows us to include interactions on the surface of
the aggregates, such as those between the surfactant and the
solvent.

Assuming a surface tension that is constant over the surface of
the aggregate (as in Ref.\ \cite{zubarev97}), the surface energy
of the system can be calculated as
$$
  E_{\rm surf}=2\pi\alpha\beta\sigma\Biggl(\frac{L}{d}\Biggr)+{\rm const},
$$
where $\sigma$ is the surface tension of the sides of the
aggregate, and the surface energy due to the ends of the aggregate
is independent of $d$. There are several ways of dealing with the
surface tension (see for example Refs.\
\cite{sudo89,flament96,halsey90}). We tried each surface tension
and found that our results were closest to experiment using the
surface energy found experimentally in Ref.\ \cite{sudo89}. Here
it was determined that the surface tension goes as $\eta^{4/5}$.
Since the surface tension includes interactions between the
particles on the surface of the aggregate, we would expect the
surface tension to saturate at large field values. To build this
saturation into our model we assume that the surface energy goes
as $M^{4/5}$. The final form of the surface energy is then
\begin{equation}
  E_{\rm surf}=2\pi\alpha\beta\Sigma\frac{L}{d} M^{4/5},
  \label{EQ:surfE}
\end{equation}
where $\Sigma$ is a constant determining the magnitude of the surface tension.

\subsection{Total Helmholtz Free Energy}

Combining these contributions, the Helmholtz free energy of the
ferrofluid system is given by
\begin{equation}
  F=E_{\rm mag}+E_{\rm surf}-TS\approx E_{\rm mag}+E_{\rm surf},
\end{equation}
with $E_{\rm mag}$ and $E_{\rm surf}$ given by Eqs.\ (\ref{EQ:magE}) and
(\ref{EQ:surfE}) respectively. Factoring out common terms and writing
the magnetization in terms of $\beta$ we obtain
\begin{equation}
  F\approx\Biggl[\Biggl(\Biggl(\frac{16}{3\pi}+\frac{6}{\beta}\Biggr)
     \frac{d}{\beta}-\frac{d^2}{\beta^2L}-\frac{6d^2}{\beta^2\sqrt{d^2+L^2}}
     \Biggr) m^2 +\frac{2\beta\Delta}{\pi}\frac{L}{d}m^{4/5}
     \Biggr]\alpha\pi^2M_S^2+{\rm const},
  \label{EQ:F}
\end{equation}
where
$$
  m=\frac{M}{M_S}=\frac{\sqrt{3}\beta^2}{2\pi}\Biggl(
    {\rm coth}\eta-\frac{1}{\eta} \Biggr),
  \;\;\Delta\equiv\frac{\Sigma}{M_S^{6/5}}.
$$
We minimize this free energy with respect to $d$
numerically using Mathematica 4.0. The values of $\beta$, L, and
$\alpha$ are fixed by the experimental conditions, so $\Delta$ is the only
free parameter in the model. The value of $M_S$ is used only to calculate
other quantities such as $\mu$ and $\phi$. $\beta$ can also be
varied slightly, but only within reasonable limits established by the value
of $\phi$ which is given for each experiment.

This free energy can be compared to the theoretical models in
Refs.\ \cite{wang94,liu95,ytreberg00}. Our new model differs from
these previous models in three significant ways. First, this model
has only one free parameter. All other parameters are specified by
experimental conditions; even this parameter could be determined
by experiment. The second difference is our treatment of the
surface energy using the results of Ref.\ \cite{sudo89}. Thirdly,
this model permits direct comparison between the magnitudes of the
magnetic energy and entropy.

\section{results of theoretical model and crossover mechanisms}

For the figures below, all parameter values except $\Delta$ are
assigned according to the experimental conditions. Then $\Delta$
is chosen so that the theoretical results give the best fit to the
experimental data. For all of the sets of data, $T=300.0$ K and
$\mu=3.0\times 10^{-16}$ erg/Oe ($\approx$10 nm magnetite
particles). For Hong {\it et al.}\ \cite{hong97} the parameter
values are: $\beta=8.0$ (calculated from $\phi=0.04$ and
$\gamma=0.71$), and $\Delta=3.09\times 10^{-4}\; {\rm
(cm^3Oe)^2/erg}$. For Wang {\it et al.}\ \cite{wang94} the
parameter values are: $\beta=4.5$ (calculated from $\phi=0.12$ and
$\gamma=0.67$), and $\Delta=2.69\times 10^{-4}\; {\rm
(cm^3Oe)^2/erg}$. As a point of interest, due to the fact that the
surface tension includes the interaction of the particles at the
surface of an aggregate, one might expect the surface tension of
an aggregate in a ferrofluid emulsion to be much smaller than that
of an aggregate in a ferrofluid. Using our parameters, we obtain
$\sigma\sim 10^{-3}\; {\rm erg/cm^2}$ for the ferrofluid emulsion
in Ref.\ \cite{liu95} and $\sigma\sim 10\; {\rm erg/cm^2}$ for the
ferrofluid in Refs.\ \cite{hong97,wang94}, confirming this
expectation.

\begin{figure}
\includegraphics{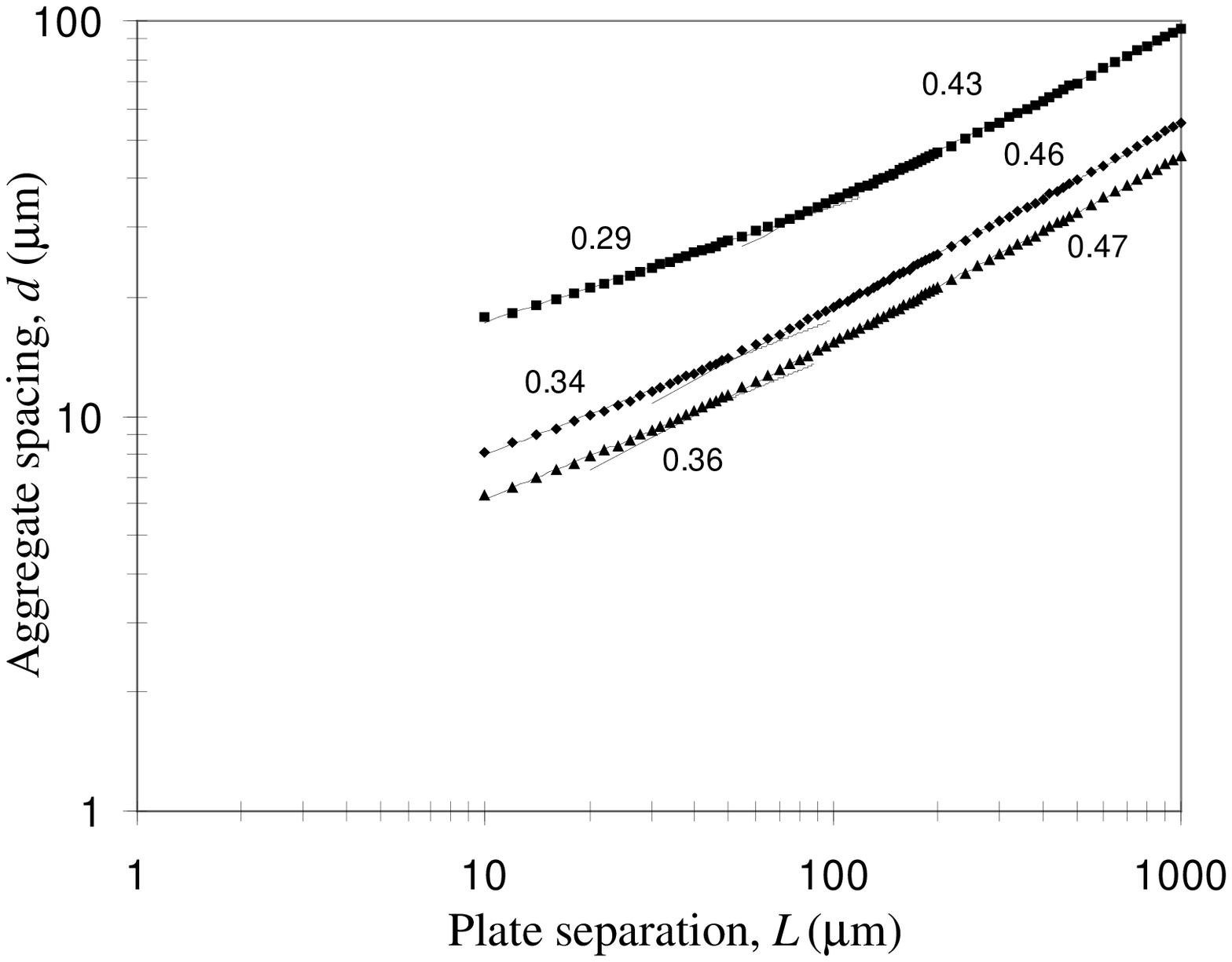}
  \caption{\label{FIG:wang94} Aggregate spacing as a function of
  plate separation for
  $H_0=100.0$ Oe (upper), $H_0=300.0$ Oe (middle)
  and $H_0=600.0$ Oe (lower), with other parameters chosen to
  model the system in Ref.\ \cite{wang94}. The exponents obtained by a power law
  fit of our numerical results are shown in the figure. }
\end{figure}

We first compare our model to the experiments in
Ref.\ \cite{wang94}, where experimental data was recorded for $d$ as a function
of $L$ for fixed $H_0=300$ Oe.
They report an exponent of 0.47, changing to an exponent of 0.67 at
$L\approx 20\;\mu$m. In Fig.\ \ref{FIG:wang94}, the results of our model are
shown as diamonds, squares and triangles for three different values of $H_0$.
The solid lines are power law fits of our numerical
results. Although, we predict lower exponents than the data
suggests, the crossover in the exponent is present. We have included results
for three values of $H_0$ to show the effect of changing $H_0$ in our model.
It is clear that increasing the field produces three significant effects:
decreasing the critical value of
$L$ where the crossover takes place, increasing the value of both exponents,
and decreasing the difference between the exponents. It is our hope that these
effects will be probed experimentally.

A closer look at our model reveals the mechanism responsible for the crossover.
In Ref.\ \cite{wang94}, the crossover in the exponent was thought to be
the result of a structural change in the aggregates for increasing $L$.
In their experiment, as $L$ is increased, the aggregates begin to lose their
circular cross sections, forming branching structures between aggregates.
A theoretical model was developed to account for these branched structures,
predicting a larger exponent as the branching occurs.
We have obtained a crossover in the exponent assuming cylindrically shaped
aggregates with no structural changes, and thus the crossover in our
model cannot be attributed to the mechanism in Ref.\ \cite{wang94}.

Examination of Eq.\ (\ref{EQ:F}) for $L\gg d$ provides a mechanism
for the crossover. In this limit, the free energy reduces to the
form
\begin{equation}
  F_{L\gg d}\approx C_0d+C_1\frac{L}{d},
\end{equation}
where $C_0$ and $C_1$ are constants with respect to $d$ and $L$.
This form of the free energy predicts
that $d\propto \sqrt{L}$. When $d \sim L$ the exponent will be less than 0.5,
with the exponent depending upon the value of $\beta$. Thus, the mechanism
for the crossover in our model is a competition between terms in the magnetic
energy. For $L\gg d$, the magnetic energy is dominated by the self-energy
of the disks and the energy due to disks in the same plane. As $L$ is
reduced, the contributions due to the disks in opposite planes become more
important, until at some critical value of $L$, they can no longer be ignored,
lowering the value of the exponent.

\begin{figure}
\includegraphics{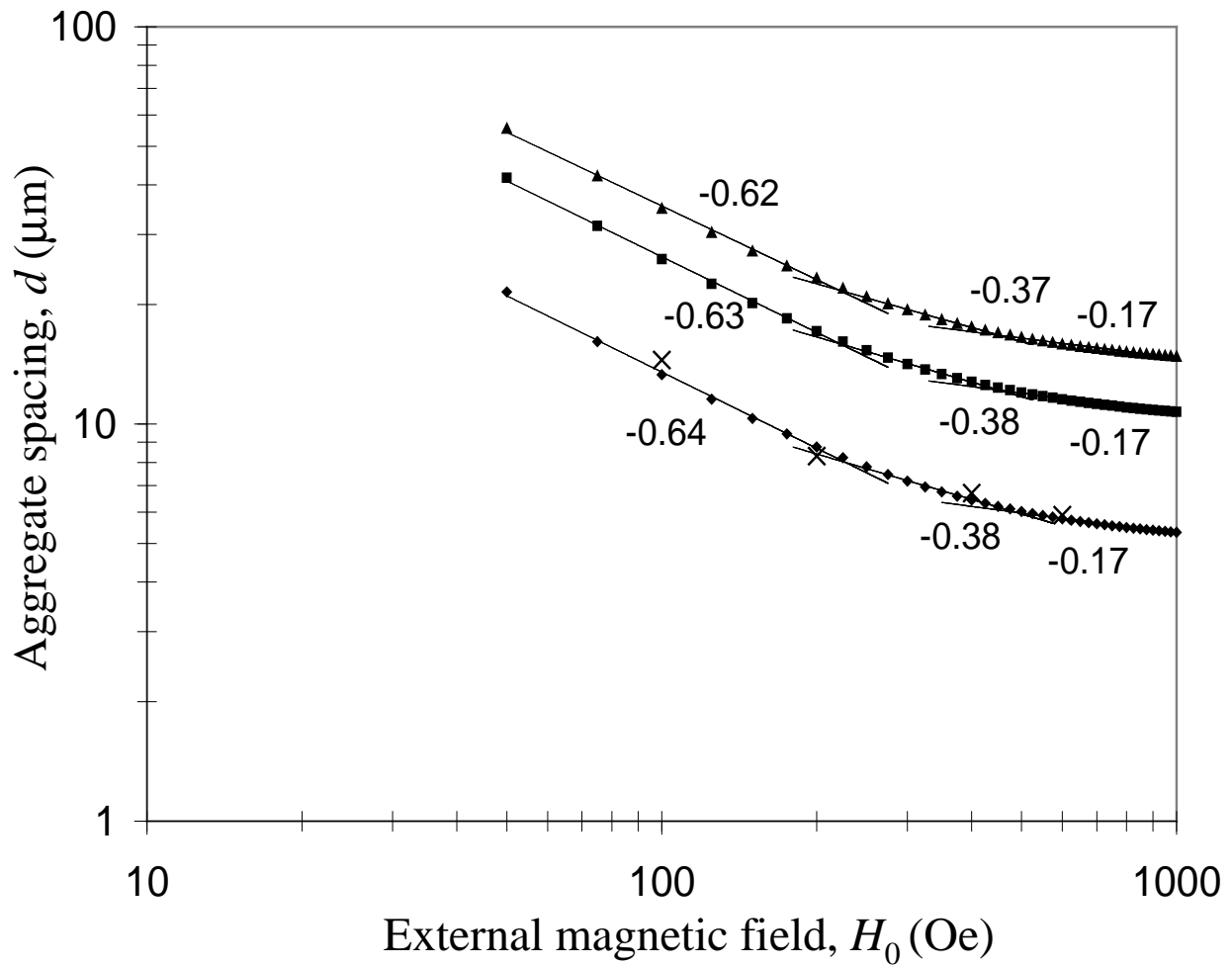}
  \caption{\label{FIG:hong97} Aggregate separation as a function of external field for
  $L=100\;\mu$m (upper), $L=50\;\mu$m (middle) and $L=10\;\mu$m (lower),
  with other parameters chosen to model the system in Ref.\
  \cite{hong97}. The exponents obtained
  by a power law fit of our numerical results are shown in the figure. }
\end{figure}

Fig.\ \ref{FIG:hong97} shows the aggregate spacing as a function of $H_0$.
The crosses represent the experimental data for $L=10.0 \;\mu$m obtained
from Ref.\ \cite{hong97}, and the diamonds, squares and triangles represent
the numerical results for three different values of $L$. The solid lines
are power law fits to our results. More experimental
data points would prove beneficial in testing the theory in this case;
however the theory agrees very well with this experimental data.
Fig.\ \ref{FIG:hong97} shows that our model predicts a crossover
in the exponent for $d$ as a function of $H_0$.
This is the first time such a prediction has been reported in the literature.
Three values of $L$ are shown to demonstrate another prediction;
the critical value of $H_0$ where the crossover occurs,
and the value of the exponents, have essentially no dependence on $L$.

From Fig.\ \ref{FIG:hong97}, it is clear that, for large $L$, the
crossover in the exponent for $d$ as a function of $H_0$ occurs
with approximately the same exponents as for small $L$. Therefore,
to understand the mechanism responsible for this crossover, it is
beneficial to look again at Eq.\ (\ref{EQ:F}) in the limit of
$L\gg d$. In this limit,
\begin{equation}
  F_{L\gg d}\approx C_2dM^2+C_3\frac{1}{d}M^{4/5},
\end{equation}
where $C_2$ and $C_3$ are constants with respect to $M$ and $d$.
This predicts that $d\propto M^{-3/5}$. Thus, for small $\eta$,
the aggregate spacing goes as $H_0^{-3/5}$, and for large $\eta$
the magnetization saturates and the aggregate spacing does not
depend upon $H_0$. For the results presented in Fig.\
\ref{FIG:hong97}, this saturation occurs $\sim$ 2000 Oe. Since the
magnetization of an aggregate is proportional to the Langevin
function, it is clear that a change in the magnetic moment of the
particles in the ferrofluid changes the value of the field where
the crossover takes place. Thus, the mechanism for the crossover
in the exponent for $d$ as a function of $H_0$ in our model is the
field dependence of the magnetization of the aggregates. This is a
direct result of the form chosen for the field dependence of the
surface and magnetic energies in our model, so experimental
studies of this crossover could be used to determine the
appropriateness of these field dependencies.

\section{conclusion}

We have shown that aggregate spacing can be determined by minimization
of the Helmholtz free energy. We have developed a simple model and
compared its predictions to experiments in Refs.\ \cite{hong97,wang94}.
Our model predicts trends as a function
of external field strength and plate separation. The results
of our model are in good agreement with experiment.

The scaling behavior of the aggregate spacing as a function of
plate separation and external field is discussed. For aggregate
spacing as a function of plate separation, there is a predicted
crossover in the value of the exponent at some critical plate
separation. This feature of our model is in agreement with Refs.\
\cite{liu95,wang94}, where similar results were reported. The
mechanism responsible for this crossover in our model is a
competition between terms in the magnetic energy.

Our model also predicts a crossover
in the exponent as a function of external field. This has not been reported in
the literature to date. The mechanism responsible for this crossover in our
model is the field dependence of the magnetization of the aggregates.
Different forms of the magnetization,
surface energy or magnetic energy would change this behavior. This result
suggests that experiments measuring this crossover could be used to probe
the forms of these quantities.

\pagebreak

\end{document}